\documentstyle[prd,aps,multicol,epsf]{revtex}

\newcommand{\U}{}

\newcommand{\f}{\frac}

\newcommand{\bb}{\bibitem}
\newcommand{\BF}{\begin{figure}\begin{center}}
\newcommand{\EF}{\end{center}\end{figure}}
\newcommand{\BE}{\begin{equation}}
\newcommand{\EE}{\end{equation}}
\newcommand{\BEA}{\begin{eqnarray}}
\newcommand{\EEA}{\end{eqnarray}}
\newcommand{\ti}{\textit}

\newcommand{\mbh}{M_{BH}}
\newcommand{\ms}{M_{\odot}}

\begin{document}
\title{Gravitational Waves from Sub-lunar Mass Primordial 
Black Hole Binaries \\
- A New Probe of Extradimensions -}
\author{Kaiki Taro Inoue$^1$ and Takahiro Tanaka$^2$}
\address{$^1$Division of Theoretical Astrophysics, National Astronomical Observatory,
Mitaka, Tokyo 181-8588, Japan
\\ $^2$Yukawa Institute for Theoretical Physics, Kyoto University, 
Kyoto 606-8502, Japan}
\date{\today}

\maketitle

\thispagestyle{empty}
\begin{abstract}

In many braneworld models, 
gravity is largely modified at \U{the} electro-weak scale 
$\sim $ 1TeV. 
In such models, primordial black holes (PBHs) 
with a lunar mass $M\sim 10^{-7}M_{\odot}$
might have been produced when the temperature of the universe was at 
$\sim$1TeV.
If \U{a} significant fraction of the dark halo of our galaxy consists of
these lunar mass PBHs, a huge number
of BH binaries will exist in our neighborhood. 
Third generation detectors 
such as EURO can detect
gravitational waves from these binaries, and  
can also determine their chirp mass. 
With a new detector designed to be sensitive at high frequency bands 
$\gtrsim 1$kHz, the existence of extradimensions
could be confirmed.
\end{abstract}

\begin{multicols}{2}
\noindent
{\it Introduction})
In recent years, there has been \U{a} great interest in 
braneworld scenarios in which the ordinary matter fields are 
confined in a four dimensional object (``brane''), 
while the gravitational field propagates in the 
higher dimensional spacetime (``bulk''). 
One motivation for the brane world scenario 
is solving the hierarchy problem between \U{the} Planck scale 
and \U{the} electro-weak scale ($\sim 1$TeV). 
The possibility of relating the Planck scale 
to the TeV scale by considering large sized extradimensions 
was pointed out~\cite{Arkani-Hamed98}. 
In the model proposed by Randall and Sundrum (RSII)~\cite{RS2} 
one of the extradimensions extends infinitely, but 
compactification of the extradimension is effectively achieved 
by the warped geometry. In this model \U{the} four dimensional Planck 
mass $M_{p}$ is related to the brane tension $\sigma$ and 
the bulk curvature length $l$ as $M^2_{p}= 4 \pi \sigma \ell^2/3$. 
Hence, keeping the scale of the brane tension at the TeV scale, 
\U{the} Planck scale can be derived by setting $\ell\approx 0.1$mm, 
although this fact does not mean that the RSII model solves the 
hierarchy problem.  

In such braneworld scenarios, the evolution of the universe 
at a temperature above $\sim 1$TeV can be dramatically altered.
For instance,  the brane universe may have had an era
of violent ``mesoscopic'' activity at scale as large as $l \sim 1$mm 
at a temperature $T\gtrsim 1$TeV~\cite{Hogan00}. 
Such a violent activity may generate 
large-amplitude fluctuations at the horizon scale $H^{-1}$. 
Then a large number of primordial black holes (PBHs) 
with lunar masses $\mbh\sim H^{-1}M_p^2\sim 10^{-7} M_{\odot}$
could be produced.
To date, PBH physics
in the context of braneworld scenarios has been 
studied~\cite{Guedens02,Majumdar03}, but none of the literature 
has shed any light on the astrophysical implication of 
considering PBHs as cold dark matter. 

Currently, there is no stringent observational 
constraint on SUb-Lunar mass Compact 
Objects (SULCOs) with mass $10^{-12}\ms \lesssim M \le10^{-7}\ms$ as 
the dark matter candidates, 
which can be either molecular clouds, small planets, or PBHs. 
Microlensing tests such as the EROS and MACHO 
collaborations ruled out the possibility that compact objects with
mass $10^{-7}M_{\odot}\lesssim M\lesssim 10^{-3}M_{\odot} $ make
up the halo dark matter in the Milky Way~\cite{Alcock98}.
On the other hand,  femtolensing tests of Gamma-ray bursts (GRBs) ruled
out the mass range $10^{-16}M_{\odot}\lesssim M\lesssim 10^{-13}M_{\odot}$
in the universe assuming that GRBs are at cosmological
distance so that the angular size of the GRB source is 
sufficiently small~\cite{Marani99}.
SULCOs also induce picolensing of GRBs, but the observational limit 
is very weak~\cite{Marani99}. \U{Dynamical constraints}
on the number of SULCOs in the dark halo 
are also less stringent~\cite{Carr99}.

In this letter, 
we consider the PBH SULCOs that have been produced by violent activity
at temperatures $\!\sim\!$TeV as the dominant constituent of the
cold dark matter.
Then there must be a huge number of 
PBHs ($\gtrsim 10^{19}$) in the dark 
halo of the Milky Way. As is discussed in the 
PBH MACHO scenario~\cite{Nakamura.etal.97}, it is natural to
expect that most of the PBHs form binaries and some of them 
coalesce emitting gravitational waves within the present age of the universe. 
From the frequency evolution in such gravitational waves, we may be able to
probe the deviation from Newtonian gravity typical 
of \U{scenarios} with infinite extradimensions. 

\noindent
{\it Formation and evolution of PBH SULCO binaries}) 
We consider the formation of PBHs in the context of gravity 
theories which have a critical energy scale at the TeV scale. 
If PBHs are formed 
at a temperature $T_c\sim \textrm{TeV}$, \U{they} are of the lunar mass,  
$\mbh \sim H^{-1}M_p^2 \sim 0.1\textrm{mm}~M_p^2 \left
(T_c/\textrm{TeV} \right )^{-2} \sim 10^{-7} M_{\odot} \left (T_c/\textrm{TeV} \right )^{-2}$. 
Here we have assumed that the number of 
effective degrees of freedom at the PBH 
formation epoch is about 100.
Although it has been argued that PBHs could undergo substantial
growth by accreting material from the cosmological background in the
high-energy regime ($\rho\gtrsim \sigma$)~\cite{Guedens02,Majumdar03}, 
such an effect is not 
relevant in our discussion because here PBHs are formed right after the 
end of the high-energy regime.  

In order to estimate the event rate of coalescence of PBH 
SULCO binaries, 
we follow the arguments in \cite{Nakamura.etal.97}. 
For simplicity, we assume that the PBH SULCOs dominate 
the dark matter, \textit{i.e.}, $\Omega_m=\Omega_{PBH}$, 
and the PBH mass spectrum is monochromatic.

At matter-radiation equality, the comoving mean separation is
$
\bar{x}=2\times10^{16} (\mbh/\ms)^{1/3}(\Omega_m h^2)^{-4/3}
\textrm{cm},
$
where $h$ is the Hubble parameter in units of 
100km $\textrm{s}^{-1} \textrm{Mpc}^{-1}$.
Henceforth, we adopt $\Omega_m h^2=0.15$ as suggested by recent observations. 
A binary with its separation $x$ decouples from the cosmic expansion 
when the local energy density 
becomes a few times larger than the energy density of the background radiation.
The tidal force from the nearest PBH at a comoving distance $y$
to the center of mass of the binary adds angular momentum to keep 
the holes from a head-on collision. 
For simplicity, we assume that 
the distribution of PBHs 
are spatially homogeneous and isotropic. 
Then the distribution of $x$ and $y$ takes the form
$p(x,y)\propto x^2 y^2 \exp({-y^3/\bar x^3})$,
which is further approximated by a uniform distribution 
$\propto x^2y^2$ in the range $x<y<\bar{x}$.  

The coalescence time $t$ due to emission of
gravitational waves can be written in terms of \U{the} 
semi-major axis $a$ and the eccentricity $e$ as 
$t=t_0 (a/a_0)^4 (1-e^2)^{7/2}$, where $t_0=10^{10}$yr and 
$a_0=2.4\times 10^{11}(\mbh/\ms)^{3/4}$cm is the initial separation of a
circular binary which coalesces in $t_0$. Thus, writing $x$ and
$y$ in terms of  $t$, $e$, and $\bar{x}$, one obtains
the distribution of the coalescing time: 
\begin{eqnarray*}
p(t) dt =\f{3}{29} \left [ \left( \f{t}{t_{\max}} \right)^{3/37}
- \left ( \f{t}{t_{\max}}\right)^{3/8}    \right ] \f{dt}{t},
\end{eqnarray*}
where $t_{\max}=t_0 ( \bar{x}/a_0 )^4$.
Let us denote the remaining time before coalescence by $T$. 
For small $T$, we can 
neglect the eccentricity $e$ because the radiation reaction 
acts to reduce it.  
Then, $T$ has a one-to-one correspondence with the semimajor axis, 
and hence with the frequency of the emitted gravitational wave $f_g$. 
The explicit relation becomes
\begin{eqnarray*}
 f_g \sim 3\times 10^3 \textrm{Hz} 
    \left ( \f{\mbh}{10^{-7} \ms} \right )^{-5/8}
    \left( T\over \textrm{5 yr}\right)^{-3/8}. 
\end{eqnarray*}
Hence, provided that $T \ll t_0$, 
the probability distribution for $\log f_g$ is obtained as 
\begin{eqnarray*}
 f_g P(f_g) & = & p(t_0+T) {dT\over d\log f_g} \cr
    & \sim & 2\times 10^{-12} 
      \left({T(f_g)\over \textrm{5yr}}\right)
         \left ( \f{\mbh}{\ms} \right )^{5/37}. 
\end{eqnarray*}
\noindent
{\it Gravitational waves from PBH SULCO binaries}) 
Now we discuss \U{observability} of gravitational waves from PBH SULCO 
binaries. 
We assume that the local dark halo density in the solar neighborhood is 
0.0079$\ms \textrm{pc}^{-3}$~\cite{Alcock00}.  
Let the total number of PBHs within the distance 
$D_s$ be $N(D_s)$. Then the distance $D_{\min}$ such that 
we can expect at least one coalescence event within $T$ at $D \le
D_{\min}$ is obtained by solving $N(D_{\min})f_g P(f_g)=1$.
Thus we have 
\begin{eqnarray*}
D_{\min}\sim  11 {\textrm{pc}} \left(\f{f_g}{100\textrm{Hz}} 
\right)^{8/9} \left(\f{\mbh}{10^{-7}\ms} \right)^{281/333}. 
\label{eq:5}
\end{eqnarray*}
After averaging over \U{the} orbital period and the orientations of 
the binary orbital plane, the 
amplitude of gravitational waves from a coalescing binary in
a circular orbit with distance $D$ at frequency $f_g$ is
$
h \sim 6\times 10^{-23}$
$ ({D}/{20 \textrm{Mpc}})^{-1}
({\mbh}/{\ms})^{5/3}  ( {f_g}/{100
\textrm{Hz}})^{2/3}. 
$
Substituting $D$ here with $D_{\rm{min}}$, we obtain 
the characteristic amplitude $h_c=h \sqrt{n} $, 
where $n=f_g \Delta T$ is the number
of cycles during the observation time $\Delta T$.
Then the wave strength of gravitational waves from a monochromatic source 
observed by using an interferometer $\tilde{h}_s=h_c/(5 f_g)^{1/2}$ is evaluated as 
\begin{eqnarray*}
\tilde{h}_s\sim 1\times 10^{-24} 
\left(\f{\mbh}{10^{-7} \ms} \right)^{{274\over 333}}
\left( \f{f_g}{100 \textrm{Hz}} \right)^{-{2\over 9}} \left( \f{\Delta
T}{ 5\textrm{yr}}\right)^{{1\over 2}}, \label{eq:hs} 
\end{eqnarray*} 
where we have taken into account the 
antenna pattern of the detector sensitivity averaged over the direction
to the binary by multiplying the factor $1/\sqrt{5}$.
The rms of the signal to noise ratio averaged 
over the source directions and orientations is given by
$\tilde{h}_s(f_g)/S^{1/2}_n(f_g)$, where 
$S^{1/2}_n(f)$ is the strain sensitivity in unit of Hz$^{-1/2}$.
Note that the above estimate for $\tilde h_s$
is valid only for frequencies 
below $f_s=3\times 10^{3}(\mbh/\ms)^{-5/8}(T/\textrm{5 yr})^{-3/8}$
for which the source frequency does not change rapidly during the 
observation time $\Delta T \ll T$.

The planned spectral noise density of the 
European Gravitational Wave Observatory (EURO) is~\cite{EURO},
\begin{eqnarray*}
S_n(f)&=&10^{-50}\times \bigl[
     \left({f/ 245{\rm Hz}}\right)^{-4} 
   + \left({f/ 360{\rm Hz}}\right)^{-2} 
\nonumber
\\
&&\qquad \qquad + 
\left({f_k/ 770{\rm Hz}}\right)(1 + f^2/f_k^2)\bigr] \textrm{Hz}^{-1},
\end{eqnarray*}
where $f_k$ 
is the knee-frequency. 
As seen from Fig.1, gravitational waves from PBH SULCO binaries 
are detectable by such an interferometer for integration time $\Delta
T \!>\!5$yr.  
Furthermore,  we can measure the chirp mass 
$M_c=(M_1 M_2)^{3/5}/(M_1+M_2)^{1/5}$ of each binary, where
$M_1$ and $M_2$ denote the masses of respective PBHs. 
This is because 
the change in frequency during the observation time 
$\Delta T$, $\Delta \nu\!=\!(df_g/d t) \Delta T$, becomes 
sufficiently large compared with the frequency resolution $\!\sim\!
1/\Delta T$ at $f_g\!\gg\! 2$Hz for $\Delta T= 5$yr. 
Thus, the existence of SULCO binaries can be confirmed by
observing gravitational waves. 

\noindent
{\it Probing extradimensions from gravitational waves}) 
\U{If PBH SULCO binaries are detected, we will have a chance to
probe the existence of extradimensions.} 
Here we focus on the RSII model as a typical example 
in which the Kaluza-Klein (KK) spectrum 
is continuous with no mass gap. 
In our present context, 
the brane tension is to be set to $\sim$(TeV)$^4$ since it 
is expected to be related to the energy scale of the Standard 
Model fields localized on the brane.  
Then, the observed four dimensional Planck 
mass is reproduced by setting 
the bulk curvature to the observationally allowed 
maximum value $\sim 0.1$mm.  
The cosmic expansion law in this model is given 
by  $H^2=8\pi (\rho+2\rho^2/\sigma)/3M_p^2$, 
where $\rho$ is the energy density of the universe. 
When $\rho\gtrsim \sigma$, 
the cosmic expansion law dramatically changes. 

Let us consider the non-relativistic
gravitational potential $V$ for a binary that consists of two point
particles with mass $M_1$ and $M_2$ on the brane. 
Let $M=M_1+M_2$ be the total mass and $\mu=M_1 M_2/M$ be
the reduced mass. The separation between the two particles is denoted by
$r$. Continuous KK modes add a correction 
to the standard Newtonian term as~\cite{GT}, 
\begin{eqnarray*}
V(r) =-\f{G \mu M}{r}\bigg [1+\f{\alpha}{(GM)^2} 
\left(\f{G M}{r}\right)^2 \bigg],~~~\alpha=\f{2 l^2}{3}, 
\end{eqnarray*}
provided that the separation $r$ 
is much larger than the AdS radius $l$.  
Therefore, the correction owing to the continuous KK
modes is second order in $\epsilon=GM/r$, i.e., 
the second post-Newtonian (2PN) order. 
In our scenario, the coefficient of this 2PN correction is $O(1)$ because 
$l$ is related to the total mass $M$ as $l/GM\sim1$. 
This correction affects the relation between the energy 
and the orbital angular velocity of the binary, and hence 
the frequency evolution of the binary is modified. 
As for the energy loss rate from the binary, 
one can show that the correction is of higher order.\footnote{
According to the AdS/CFT correspondence, the 
energy loss owing to geometrical particle
creation in the RSII model is $30 \pi l^2/G$ times
as large as that for a single conformally coupled scalar field~\cite{Duff00}.
Using the result for 
a scalar field~\cite{jaume}, one can easily 
find that the correction is suppressed by \U{a} factor $f^2 \ell^2$, 
and hence is of 3PN order.}
 
Fourier components of gravitational waves from
\U{an} inspiralling binary in a quasi-circular orbit are given by 
\begin{eqnarray*}
\tilde{h}(f)=\f{Q}{D} (G M_c)^{5/6}f^{-7/6} \exp(i \Psi(f)),
\end{eqnarray*} 
where $Q$ is the factor
depending on the direction and the orientation of the binary 
and the detector antenna pattern.
Keeping the second order term in $\epsilon$, the phase $\Psi(f)$
can be written as 
\begin{eqnarray*}
\Psi(f)&=&2 \pi f t_c-\phi_c-\pi/4+\gamma(f)(\pi G M_c f)^{-5/3},
\nonumber
\\
\gamma(f)&=&
\f{3}{128}\Biggl (1+ \left( \f{3715}{756}+\f{55}{9}\eta
\right)(\pi G M f)^{2/3}
\nonumber
\\
&+&(4 \beta -16 \pi) (\pi G M f)+\bigg ( \f{15293365}{508032}+\f{27145}{504}\eta
\nonumber
\\
&+& \f{3085}{72}\eta^2 
-\f{100 \alpha }{(GM)^2}\bigg ) (\pi G M f)^{4/3}
                          \Biggr ),
\end{eqnarray*}
where $\eta=\mu/M$, $t_c$ is the coalescence time, $\beta$ denotes the
spin parameter, and $\phi_c$ determines the phase. 
The KK mode correction appears in the last term in $\gamma(f)$. 
Here we have neglected the spin dependent term at the 2PN order. 
One may think that the KK modes contribution 
is indistinguishable from this effect. 
However, for \U{binaries} whose spin parameter at the 
1.5PN order $\beta$ is measured to be small, 
one can neglect the spin effect at the 2PN order. 
Hence, the effect of extradimensions can be observed 
as an anomalous ``2PN'' correction.

\begin{figure}
\centerline{\epsfxsize8.5cm \epsfbox{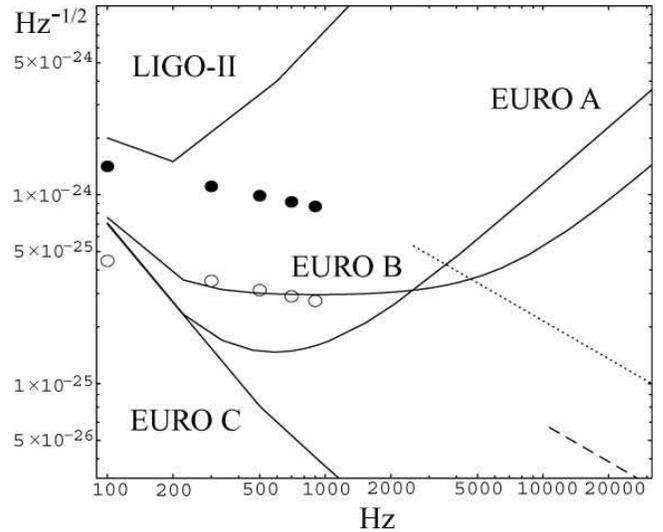}
}
\caption{Sensitivity curves $\tilde{h}_n(f)=S^{1/2}_n(f)$ 
for LIGO-II and EURO with different specifications, (A,B,C)
$f_k=10^3$Hz, $f_k=6.5\times 10^3$Hz, and the xylophone type,
 respectively.
Disks and circles denote the wave strength $\tilde{h}_s$ 
for the nearest PBH SULCO binary with mass $\mbh=10^{-7}\ms$ 
for the observational time $\Delta T=10, 1$yr, respectively.
Dotted lines and dashed lines 
represent the evolution of the amplitude $Q(G M_c)^{5/6}f^{-7/6}/D_{min}$
for the nearest PBH SULCO binary with $\mbh=10^{-7}\ms$, 
$\mbh=10^{-8}\ms$, respectively. The endpoint  
of each line in the left-upper side corresponds to the remaining time 
$T=10$yr before coalescence. }
\label{fig:1}
\end{figure}

Let us evaluate how accurately we can determine the
AdS radius $l$, using third generation detectors such as EURO.
We set $f_k=6.5\times10^{3}$Hz in the following analysis.
We assume that a PBH SULCO binary with lunar mass $M_1=M_2=10^{-7}\ms$ 
is in a quasi-circular orbit. With AdS radius $l=M_p^{-2} M=0.3$mm,
a binary that emits gravitational waves at a frequency 
$f=2.5\times10^3$Hz coalesces in 10 yr. The distance to the 
nearest binary is estimated as $D_{\min}\sim 3\times
10^2$pc, which yields the amplitude $\tilde{h}=Q (G M_c)^{5/6}D^{-1}f^{-2/3}
 \sim 1\times 10^{-24}$.

In the matched filter analysis up to 2PN 
order for the phase $\Psi(f)$, 
we marginalize seven parameters $\{\lambda_i \}\!=\!
\{\ln A, t_c, \phi_c, \ln M_c, 
\ln \mu, \beta, \ln l\}$.  
The error of each parameter can be estimated by a quadratic fitting of the
likelihood function assuming uncorrelated Gaussian noise~\cite{Cutler94}.  
Using the EURO detector, we find that \U{the nearest} 
PBH SULCO binary is expected to be detected 
with a signal-to-noise ratio $S/N=3$. 
The relative error in $l$ 
is $\Delta l/l=3\times10^{-1}$ 
whereas $\Delta t_c =2 \times 10^{-4}\textrm{sec}, 
\Delta \phi_c=5\times 10^2\textrm{rad}, \Delta M_c/M_c=7\times10^{-11}, \Delta
\mu/\mu=7\times10^{-5}$ and $\Delta \beta=4\times 10^{-2}$. 
Although the $S/N$ is not sufficiently large,  
the AdS radius $l$ can be determined within $\!\sim\!30$ percent error.  
To gain a large $S/N$ value, we need a detector that is highly sensitive
at high frequency bands ($>$1kHz), say, a ``xylophone'' 
type detector~\cite{EURO} that consists of 
several narrow-band interferometers.

\noindent
{\it{Summary and Discussions}})
In this letter, we discussed \U{observability} of 
gravitational waves from the nearest coalescing PBH SULCO binary.  
If \U{PBH SULCOs} 
with mass $\mbh\!\sim\!10^{-7}\ms$ constitute the cold dark matter,
the third generation interferometer, EURO, has sufficient 
sensitivity to confirm their existence. 
For the case that the PBH formation is related to the braneworld scenario, 
we further discussed \U{observability} of the imprint of 
extradimensions. 
We found that the sensitivity of EURO is 
marginal to discriminate the RSII model. 
However, once we discover the PBH SULCO binaries, we would be able to 
construct a new detector that is suitable for probing the extradimensions.

Although we have investigated the RSII model as an example, 
our estimate of \U{observability} is not very restrictive to this model. 
In the RSII model, the deviation from general relativity 
starts at the 2PN order. 
In other models 
the deviation may appear at lower PN order.  
Even in that case, the correlation with other parameters such as
$M_c, \mu$ and $\beta$ renders new physics 
unveiled up to the 2PN order. 
Therefore, it is plausible to conclude that our result is generic to
a rather wide class of models. 

Because the number of PBH SULCOs in the dark halo
is huge, one might expect that the  
radiation from gas accretion to the PBHs
is detectable\cite{Fujita98}. However, as we shall see,
the chance of detection is hopelessly small.
Let us consider that a lunar mass $\mbh\!=\!10^{-7}\ms$ BH is moving with velocity 
$V$ with respect to the surrounding gas.
The accretion rate from the surrounding gas
can be approximated as $\dot{m}\!\equiv \!\dot{M}_{BH}/\dot{M}_{Edd}\sim
5\times10^{-4}(\mbh/\ms)(n_g/10^2\textrm{cm}^{-3})(10\textrm{km} s^{-1}/V)^{3}$
where $n_g$ is the number density of the surrounding gas, and the 
$\dot{M}_{Edd}$ is the Eddington accretion rate with unit efficiency.
Adopting the thin-disk model\cite{Ipser77}, the total luminosity from the disk 
is estimated, at most as $L\!\sim\!G M_{BH}\dot{M}_{BH}/(2 R_{min}) $ where $R_{min}$
is the innermost accretion radius. Assuming black-body 
radiation from the disk, $\dot{m}\!=\!1$ and $R_{min}\!=\!3 R_{Schw}$
where $R_{Schw}$ is the Schwarzschild radius, 
the temperature becomes 
$T\!\sim\!1\times10^7 K (\mbh/\ms)^{-1/4}$.
Thus smaller mass BHs have a higher temperature. 
Now consider lunar mass PBHs moving in the Orion Nebulae at 400 pc
away. Assuming that the dark matter density in the solar neighborhood 
is 0.0079$\ms \textrm{pc}^{-3}$\cite{Alcock00}, the total number of 
lunar mass BHs is just $\sim 8\times10^7$ in 
the molecular cloud if its extention is $10^3\textrm{pc}^{3}$. 
Provided that the velocity distribution of the PBHs is Maxwellian, 
and the number density of the molecular gas is 
$n_g\!=\!10^2\textrm{cm}^{-3}$, we find that the 
expected total luminosity is $L\!\sim\!2\times 10^{24}\textrm{ergs/s}$.
and the peak frequency is $f\sim 10^{18}$Hz. 
On the other hand, the sensitivity of the 
 X-ray telescopes such as the \ti{Advanced X-Ray Astrophysics Facility}
(AXAF) is about $\sim 10^{29}\textrm{ergs/s}$ for a source at 400pc
at $f\sim 10^{18}$Hz.
Thus we cannot expect any X-ray visible BHs in the Orion molecular cloud. 

Based on the AdS/CFT correspondence, 
some authors have conjectured that 
the evaporation rate of BHs in the RSII model
is greatly increased~\cite{Tanaka02,Emparan02}. 
The lifetime of massive \U{BHs} in the RSII model is estimated as 
$\tau\sim 10^5 (\mbh/10\ms)^3 (0.1\textrm{mm}/l)^2$yr 
for a BH with a mass $\mbh\gg l M_p^2$~\cite{Tanaka02,Emparan02}. 
Thus, most astrophysical BHs with moderate
mass $10\ms \lesssim \mbh< 10^3 \ms$ would have evaporated by now, 
leaving remnants with a small mass $\sim l M_p^2$. 
If this should be the case, we would be able to 
observe gravitational waves from \U{BH} SULCO binaries that
are not primordial ones, though the observational evidence of 
stability of stellar mass BHs would 
give a stringent constraint on the size of the AdS radius $l$. 
\\
\indent
We would like to thank K. Nakamura for useful discussions. 
We also thank W. Naylor for his careful reading of the mansript. 
This work was supported in part by Grant-in-Aid for 
Scientific Research Fund (Nos.11367, 14740165 and 14047212). 

\end{multicols}

\end{document}